# Effect of RF Sputtering Process Parameters on Silicon Nitride Thin Film Deposition

Sachin S Bharadwaj[a], Shivaraj B W[a*], GR Rajkumar[a], M Krishna[a],

[a]Department of Mechanical Engineering, R V College of Engineering, Bangalore, 560059, India

**Abstract**

The objective of this work was to study the RF sputtering process parameters optimisation for deposition of Silicon Nitride thin films. The process parameters chosen to be varied were deposition power, deposition duration, flow rate of argon and flow rate of nitrogen. The parameters were varied at three levels according to Taguchi L$_9$ orthogonal array. Surface topology, film composition, coating thickness, coating resistivity and refractive index were determined using SEM, XRD, profilometer, Semiconductor device analyser and UV spectrometer respectively. The measured film thickness values ranged from 127.8nm to 908.3nm with deposition rate varying from 1.47nm/min to a maximum value of 10.1nm/min. The resistivity of the film varied between $1.53 \times 10^{13}\Omega$m to $7.85 \times 10^{13}\Omega$m. Refractive index of the film was calculated to be between 1.84 to 2.08. From the results, it was seen that film properties tend to be poor when there is no nitrogen flow and tend to improve with small input of nitrogen. Also, SEM images indicated amorphous structure of silicon nitride which was confirmed by XRD pattern.

*Keywords*: Silicon Nitride, Design of experiments, Surface topology, Refractive index, Resistivity.

## 1. Introduction

In modern technology, the role of dielectric thin films in the fabrication of integrated microelectronic circuits is vital. Applications of these dielectric thin films include dielectric insulation, diffusion masking, surface passivation, radiation resistance and hermetic sealing. These vital applications of the films entail stringent requirements on the properties of these films. Dielectric layers in thin film devices can be deposited by various methods such as vacuum deposition, sputtering, vapor deposition and anodizing [1]. To date, silicon dioxide has been used almost exclusively as a dielectric material for microelectronic applications for a few reasons because ease of preparation, its physical, chemical and dielectric properties are well understood, it is well compatible and forms good interface[2]. There are various dielectric materials such as silicon monoxide, silicon dioxide, rubidium bromide .The inorganic dielectric materials that are now being used or are under investigation for future high-K dielectric materials are silicon monoxide (SiO, K~5.0), silicon dioxide (SiO2, K~3.9), silicon nitride (SiO3N4, K~6), However silicon nitride has two distinct advantages over silicon dioxide because it provides a much better barrier to moisture and sodium ions, which is major detrimental factors in any thin film device and as a conventional nitride it is opaque to UV.

Silicon nitride films are commonly prepared by PVD, PECVD and CVD techniques. These films contain relatively high amounts of hydrogen, which can lead to a degradation of films in subsequent high temperature processing steps [3]. Therefore, sputtering is an interesting thin film deposition technique for all silicon nitride applications where low processing temperatures are desired, low hydrogen contents in the films are required because no hydrogen content is obtained or the stoichiometry of the films should be controlled for example–to obtain high hardness of the film [4].Silicon nitride has the strongest covalent bond properties next to silicon carbide. The Young's modulus of silicon nitride thin film is higher than that of silicon and its intrinsic stress can be controlled by the specifics of the deposition process [5]. Batan et al. [6] deposited SIN films using DC reactive magnetron sputtering. XPS spectra indicated that Si$_3$N$_4$ was obtained for a molar fraction of nitrogen larger than 0.1, and also indicated that the sputtered SIN films were uncontaminated with oxygen and carbon. Serikawa and Okamoto [7] deposited SIN films of 100–200 nm thickness at 200°C by RF planar magnetron system. SIN films have been synthesized with 1.7–200 nm/min with an etching rate in buffered hydrofluoric acid, 2.97-2.28g/cm$^3$ of density, 1.97-1.82 refractive index and compressive intrinsic stress of 5 x$10^9$-1 dyne/cm$^2$ with breakdown voltage of 5.5x$10^6$-3.5x$10^6$V/cm. Kim and Chung [8] deposited SIN thin films by RF magnetron sputtering, it was observed with the application of a substrate bias of −50 Voptical band gap of the SIN films varied from 4.85 to 4.39 eV.

Herman J. Stein [9] investigated the characteristics of SIN films by LPCVD and the SiN films were deposited to a thickness of ~500Å on a 100 mm diameter crucible grown n-type Si wafers. The deposition was carried out at 750°C with 9:1 ammonia: dichlorosilane ratio and refractive index was 1.98 at 6328Å. S. M. Hu et al [10] studied the effect of process variables on the properties and composition of SIN thin films ranging from 500Å to 7500 Å by D.C and R.F. Sputtering with the mixture of argon and nitrogen was used in the atmosphere. SIN films grown above 10 % nitrogen were transparent and insulating, while those below 5% were semiconducting with a metallic lustre. The dielectric constant of SIN was in range of 6.2 to 8.5. M. Lattemann et al [11] deposited SIN thin films by RF sputtering ,films were deposited at a power of 300W at a distance of 170 mm between the substrate and target. Substrate temperatures were varied as 120, 350, 525 and 700 °C. The structure of the films was characterized by XRD, FTIR and Raman spectroscopy. O. Debieu et al [12] investigated the relationship between Si content, optical and structural properties of O- and H- free SIN thin films deposited by magnetron sputtering with the varying thickness of films from 100 to 200 nm. The compositions of the film were obtained by Rutherford Back Scattering (RBS), XRD, FTIR, Raman Spectroscopy and ellipsometry. G. Eisenstein et al [13] deposited SIN films by sputtering with a deposition rate of (~75Å/min) allows film thickness control to within ±50Å with a control over refractive index which can be achieved by adjusting partial pressure of nitrogen in the plasma. V. Bhatt et al [14] deposited SIN films by R.F. sputtering at 260°C . $Si_3N_4$ target was used in argon ambient, pressure was varied from 5 to 20 mTorr, power from 100 to 300 W. The deposition rate was found to vary from 30Å/min (100W, 5mTorr) to 190 Å/min (300W, 20mTorr), refractive index was in the range of 1.84-2.04.

Review of literature [1-14] concentrates on higher temperature of deposition of the silicon nitride thin films. However, research on a lower temperature of deposition is scarce. The scope of this research work was to investigate sputtering process parameters for lower temperature of deposition, in order to obtain a high quality film capable of performing as a superior dielectric. The objective of this work was to perform a detailed study of the RF sputtering process parameters for low temperature deposition of Silicon Nitride thin films and characterize. Based on the observed behavior of the film and its performance as a dielectric, optimum process parameters for deposition of the film can be found and validated from its reproducibility.

2. *Materials*

*2.1. Materials and Equipment used*

Silicon Nitride($Si_3N_4$) target of 75mm in diameter, 5mm in thickness and was of 99.5 % pure. Float glass with a composition of 73% $SiO_2$, 14% - $Na_2O$, 9% - CaO and 4% of MgO was selected as substrate. Silicon nitride was deposited on square substrate of 30*30mm size. A semiautomatic RF sputter coater was employed to coat silicon nitride. The ranges selected were, power between 100-200W, time of deposition between 40 – 90 minutes, argon flow rate between 30-50 sccm, nitrogen flow rate between 0-30 sccm.

X- rays directed on to the coated samples of dimension 30x30mm using X-Ray diffraction equipment. For a range of angles, the reflected rays were captured and the corresponding intensity of the reflection V/s angle of incidence plots were obtained. The peaks obtained were examined for composition analysis. The equipment used for RF sputtering was Solar Metalization unit. The semi-automatic unit has a maximum RF power supply of 200W, DC power supply of 2 Kw and two numbers of 3 inch flexible magnetrons. The maximum temperature that can be achieved by the machine is 275°C. The three gauges used to monitor pressure at various locations are Pirani, penning and capacitance monometer. Pirani gauge measures pressure in the range of 1 millibar to $10^{-3}$ mbar, penning gauge in the range of $10^{-2}$ to $10^{-6}$ millibar and to measure pressure between 10 to 10$^{-4}$ mbar capacitance manometers is used. Scanning electron microscope( SEM) images were used to study the surface topology and composition by Hitachi SU 1500 having magnification of 5 Lakh X. A thin layer of gold was sputtered onto the samples in order to provide a conducting layer over the dielectric silicon nitride thin film. Sputtering was done for a period of 160s. On completion of sputter deposition on silicon nitride, the chamber pressure gradually decreases to atmospheric pressure, simultaneously the temperature in the chamber decreases to around 60˚C. The samples were loaded onto the machine and on focussing the SEM images revealed that the coating was amorphous in nature. The samples are then wrapped in butter paper to prevent contamination and are then characterised for basic and functional properties. Images at different levels of magnification of each sample were taken.

Profilometer was used for measurement of film thickness. A step was created on substrate through deposition of silicon nitride using kepton tape. A stylus measures the height of the film at various points by fixing the step as zero (measured with respect to substrate surface). The average of the values were taken as film thickness and for different morphological and surface characterisations. UV Visible spectroscope was used to obtain the transmittance graph of percentage transmittance vs wavelength (nm). For resistance measurement, Semiconductor Device Analyser called Agilent B1500A s was made use of with the probes are placed at fixed distance in the range of few millimetres and the current is passed between them in the order of few nano maters to obtain Current v/s Voltage plots. The resistance is measured for a series of voltage value and the average is taken.

*2.2 Preparation of Setup Parameters*

For better properties the samples were annealed at 400˚C for a time period of one hou before characterization. . The rise in temperature was linear and the rate was set at 10$^0$C per minute. After an hour the heater was switched off and samples were cooled inside the chamber.
The experiments were carried out as per design of experiments (DOE), based on Taguchi technique. Present work has four factors viz. RF power(p), deposition time(T),argon flow rate(Ar) and Nitrogen flow rate (N) along with their three levels. Film resistivity and thickness are the response parameters. If all the possible combination to be considered, the number of experiments would be 81 which are tedious in terms of time and cost. Proposed DOE reduces to nine experiments (L9). However, L9 orthogonal array assumes that there are interaction between the selected parameters. The factors and their levels were assigned as per requirements of selected L9 orthogonal array in Minitab software and the experimental plan was given in Table 1 and 2.

Table 1 Factors and levels chosen for bench marking

| Factors | Unit | Level 1 | Level 2 | Level 3 |
| --- | --- | --- | --- | --- |
| Power | Watt | 100 | 150 | 200 |
| Time | Min | 40 | 60 | 90 |
| Argon flow rate | sccm | 30 | 40 | 50 |
| Nitrogen | sccm | 0 | 20 | 30 |

Table 2 L9 Experimental plan using L9 orthogonal array

| Expt. No. | Power in watt | Time in mins. | Organ flow rate In sccm | Nitrogen flow rate in sccm |
|---|---|---|---|---|
| 1 | 100 | 40 | 30 | 0 |
| 2 | 100 | 60 | 40 | 20 |
| 3 | 100 | 90 | 50 | 30 |
| 4 | 150 | 40 | 40 | 30 |
| 5 | 150 | 60 | 50 | 0 |
| 6 | 150 | 90 | 30 | 20 |
| 7 | 200 | 40 | 50 | 20 |
| 8 | 200 | 60 | 30 | 30 |
| 9 | 200 | 90 | 40 | 0 |

## 3. *Results and Discussion*

The grain size and structure of samples could be studied. The surface morphologies of the SiN films showed that the grains were very tiny in size with no well-defined grain boundaries. The cubic shaped SiN films concealed the glass substrate. The granule by some means formed condensed morphology structure over the substrate and was densely packed. The amorphous structure pertaining to experiment no.9 being the best for the power at 200W, time being 90 minutes, argon flow rate equal to 40sccm with no nitrogen supply. One can observe that the nanometers structure of the surface is uniform and continuous. The SEM for experiment 6 images showed void spaces for films produced for power at 150W, argon flow rate at 30sccm and nitrogen flow rate of 20sccm. which has a detrimental effect on the conductivity and hence performance of the films. SiN deposited at high flow rate ( 30sccm) was found to be electrically, morphogically stable than ones deposited at low nitrogen flow rates. All the SEM images are secondary electrons images.

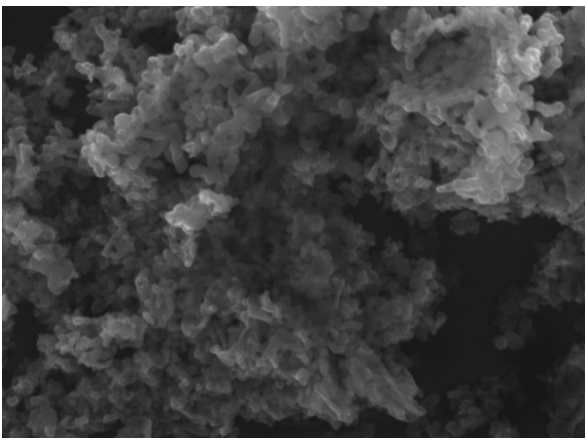

**Fig. 1 SEM image of Experiment 3**

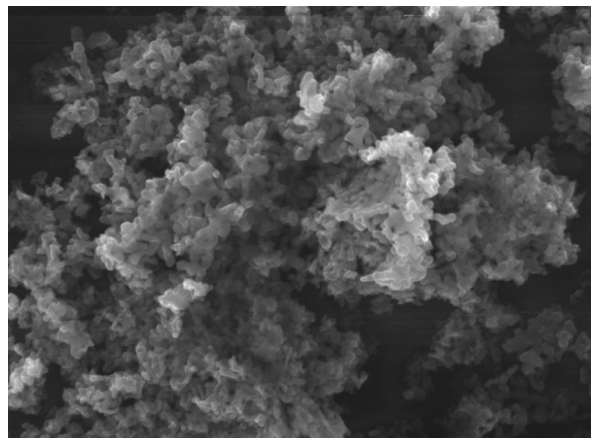

**Fig. 2 SEM image of Experiment 9**

On analysis of XRD plot (Fig.3), an increase and then a gradual decrease in diffraction peaks is seen which indicates the formation of thin film which is amorphous in nature unlike sharp increase which denotes a crystalline structure. The result plots suggest that silicon nitride indeed be amorphous in nature.

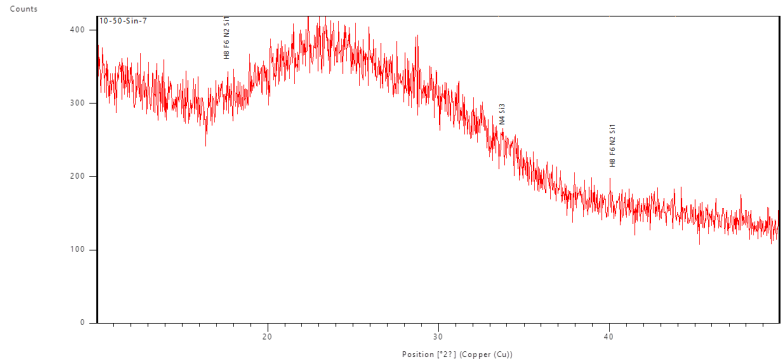

**Fig.3 XRD plot for Experiment 9**

## 3.1 Resistivity studies

The resistivity was calculated using the formula :

$$Resistivity = \frac{Resistance \; x \; Thickness \; of \; film \; in \; cm \; x \; Electrode \; Length}{Electrode \; Gap}$$

Changes in resistivity as a function of various parameters are shown main effect plot Fig.4 and in table 3. The resistivity of expt 9 where there is no nitrogen flow show is lowest and it is highest when there is nitrogen flow rate of 30sccm. It implies that resistivity increase with increase in nitrogen flow rate, which is evident from Fig.4. However resistivity reduces when power decrees from 150W to 200W to an extent of 60%. The resistivity drop with increasing in annealing time from 40 to 60 sec while with further increase in annealing time the resistivity increase. It can be clearly observed that the resistivity is proportional to argon and nitrogen flow rate. The highest resistivity of 7.85x x$10^{13}$Ωm was obtained for 100W power, 90 mins time, with Argon and nitrogen flow rate of 50sccm and 30sccm respectively. The lowest resistivity was obtained when there is no nitrogen flow. The resistivity value of experiment 3 was maximum at 7.75x$10^{13}$Ωm.

Higher nitrogen content (>30%) results in diffusion of excess nitrogen to surface and subsequent desorption, nitrogen precipitation in grain boundaries also beneficial to achieve low resistivity, making the films more suitable for microelectronic application. The nitrogen rich films showed high resistivity demonstrating better mechanical stability.

66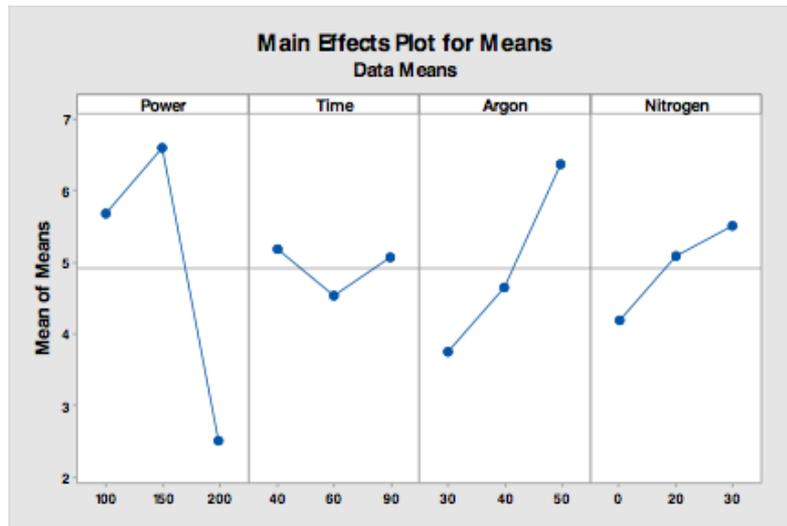

**Fig.4. Variation of resistivity with variation in parameters**

Table 3: Resistivity values for selected parameters and levels

| Trial no | Power (P) W | Time (T) min | Argon (Ar) sccm | Nitrogen (N) sccm | Thickness A° | Resistance (MΩ) | Resistivity $(\times 10^{13})\Omega$(Ωm) |
|---|---|---|---|---|---|---|---|
| 1 | 100 | 40 | 30 | 0 | 1278 | 56.89 | 4.006 |
| 2 | 100 | 60 | 40 | 20 | 1296 | 74.36 | 5.16 |
| 3 | 100 | 90 | 50 | 30 | 1325 | 115.76 | 7.85 |
| 4 | 150 | 40 | 40 | 30 | 1381 | 110.04 | 7.16 |
| 5 | 150 | 60 | 50 | 0 | 1402 | 107.96 | 6.92 |
| 6 | 150 | 90 | 30 | 20 | 1486 | 94.52 | 5.72 |
| 7 | 200 | 40 | 50 | 20 | 1547 | 75.06 | 4.36 |
| 8 | 200 | 60 | 30 | 30 | 7062 | 120.81 | 1.53 |
| 9 | 200 | 90 | 40 | 0 | 9083 | 163.68 | 1.62 |

*3.2 Profilometer Results*

We here have again taken 3 sets of data points for 100, 150 and 200W .The main effects plot for means related to thickness (Table 4) is given in Figure 2. We see that the thickness has a significant increase from 150 to 200W power. Similarly we observe a highest thickness for 90 min time. Whereas the thickness decreases for 50 sccm Argon and has a minimum for 20 sccm Nitrogen. A maximum thickness of 9083 ˚A was obtained from Expt. 9 . The parameters corresponding to Expt. 9 are power of 200W, time of 90 minutes with Argon flow rate being 40 sccm with no Nitrogen involved.

Table 4: Thickness Results

| Trial no | Power (P) W | Time (T) min | Argon (Ar) sccm | Nitrogen (N) sccm | Thickness A° |
|---|---|---|---|---|---|
| 1 | 100 | 40 | 30 | 0 | 1278 |
| 2 | 100 | 60 | 40 | 20 | 1296 |
| 3 | 100 | 90 | 50 | 30 | 1325 |
| 4 | 150 | 40 | 40 | 30 | 1381 |
| 5 | 150 | 60 | 50 | 0 | 1402 |
| 6 | 150 | 90 | 30 | 20 | 1486 |
| 7 | 200 | 40 | 50 | 20 | 1547 |
| 8 | 200 | 60 | 30 | 30 | 7062 |
| 9 | 200 | 90 | 40 | 0 | 9083 |

## 3.3 UV-VIS-Spectrometer results

Transmittance graph of percentage transmittance vs wave length was used to calculate the refractive index of the thin film using the formula

$$Refractive\ index,\ n = \frac{1 + \sqrt{1 - T_{max}^2}}{T_{max}}$$

The second line in Fig. 5 indicates the %T of silicon nitride thin film. Maximum %T value of 78 occurs at a wave length of 600nm. This graph also suggests a steep increase in %T value in initial stages (between 250nm to 400nm) and for wave length beyond 400nm there is not much change in the value. Substituting the maximum value of %T, refractive index turns out to be 2.08. This value is in close agreement to the theoretical refractive index of thin film silicon nitride which is 2.06. Raman spectroscopy was used to study the microstructure of the deposited silicon nitride. Fig. 6 shows the typical Raman shift for silicon nitride is around 550/cm

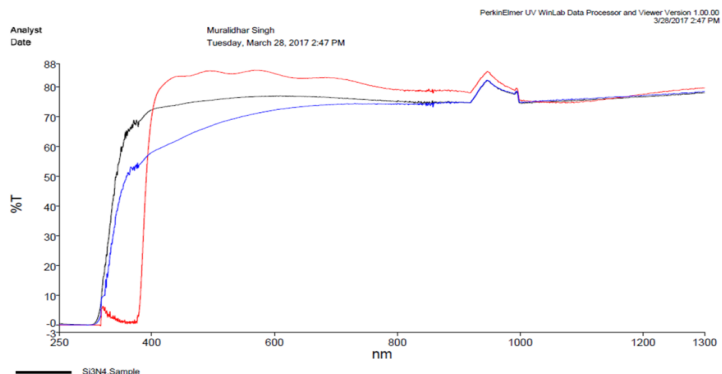

**Fig.5. %Transmittance Graph of silicon nitride.**

The graph also suggests a steep increase in %T value in initial stages (wavelength between 250 nm to 750 nm) and for wavelengths between 750 nm and 1700 nm there is a gradual increase in %T value and beyond 1700 there isn't much change in the value (graph becomes a straight line parallel to X axis). Substituting the maximum value of %T, refractive index turns out to be 1.84.

*3.4 Raman spectroscopy results*

Raman Spectroscopy was used to study the microstructure of the deposited silicon nitride. Figure 5.9 shows the typical Raman shift for silicon nitride is around 550 cm$^{-1}$. In our trials, we conducted Raman Spectroscopy for batches 5 and 6 and the following results were obtained. As we can see from Fig. 6 the Raman shift was for the samples tested is around 560 cm$^{-1}$, which confirms the presence of silicon nitride in the films deposited.

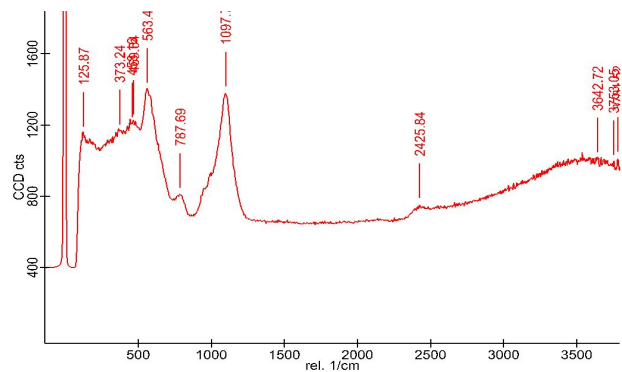

**Fig. 6: Raman shift**

*4. Conclusion*

The RF sputtering process parameters were investigated by employing Taguchi method for design of experiments which involves variation of parameters and conducting trials of certain combinations. First a basic characterization was performed to understand the nature of the film and then a functional characterization was performed to evaluate the film's ability to be used as a dielectric layer. At this low temperature of deposition, the following conclusions have been drawn.

- The parameters which gave optimum properties were power of 200W, time of 90 minutes, and argon flow rate of 40sccm. The coating thickness obtained was 908.3nm, refractive index 2.06 and resistivity of $1.53 \times 10^{13}$ was obtained.
- Low temperature of deposition yields amorphous silicon films which have low reproducibility.
- An increase in the time of deposition yields a more uniform coating of Silicon Nitride, as indicated by the SEM images, also yields improved dielectric properties such as resistivity and dielectric constant and   refractive index.

- Film properties tend to be poor when the Nitrogen flow rate is zero and they tend to improve with small input of Nitrogen.